\documentclass[superscriptaddress,prb,twocolumn,tightenlines,balancelastpage,10pt,a4paper]{revtex4}
\usepackage{bbm}
\usepackage{bm}
\usepackage{amsfonts}
\usepackage{amssymb}
\usepackage{graphicx}
\usepackage{subfigure}
\usepackage{savesym}
\usepackage{amsmath}
\usepackage{txfonts}
\usepackage{multirow}
\usepackage{epstopdf}
\usepackage{soul} 
\usepackage{color}

 \savesymbol{iint}
\restoresymbol{TXF}{iint}

\newcommand{\ket}[1]{\vert#1\rangle}
\newcommand{\bra}[1]{\langle#1\vert}

\newcommand{\red}[1] {\textcolor{black}{#1}}
\newcommand{\blue}[1] {\textcolor{black}{#1}}

\begin{document}

\author{Xiao-Qi Zhou\footnote[1]{These authors contributed equally}$^{,\,}$\footnote[2]{Email: Xiaoqi.Zhou@bristol.ac.uk}$^{,\,1}$,
Hugo Cable$^{*,}$ \footnote[3]{Email: Hugo.Cable@bristol.ac.uk}$^{,\,2,\,1}$,
Rebecca Whittaker$^*$$^{,\,1}$,
Peter Shadbolt$^1$,
Jeremy L. O'Brien$^1$,
Jonathan C. F. Matthews\footnote[4]{Email: Jonathan.Matthews@bristol.ac.uk}$^{,\,}$}
\affiliation
{
Centre for Quantum Photonics, H. H. Wills Physics Laboratory
and Department of Electrical and Electronic Engineering, University
of Bristol, Merchant Venturers Building, Woodland Road,
Bristol BS8 1UB, UK.\\
$^\textrm{\textit{2 }}$Centre for Quantum Technologies, National University of Singapore, 3 Science Drive 2, Singapore 117543
}

\title{{{\color{black}{Quantum-enhanced tomography of unitary processes}}}\vspace{-6pt}}

\maketitle

\textbf{A fundamental task in photonics is to characterise an unknown optical process, defined by  properties such as birefringence, spectral response, thickness and flatness. Amongst many ways to achieve this, single-photon probes can be used in a method called quantum process tomography~(QPT). Furthermore, QPT is an essential method in determining
how a process acts on quantum mechanical states.
For example for quantum technology, QPT is used to characterise multi-qubit processors~\cite{ob-prl-93-080502} and quantum communication channels~\cite{wa-nphot-7-387}; across quantum physics QPT of some form is often the first experimental investigation of a
new physical process, as shown in the recent research into coherent transport in biological mechanisms~\cite{yu-pnas-108-17615}.
%
%
%
However, the precision of QPT is limited by the fact that measurements with single-particle probes are subject to unavoidable shot noise---this holds for both single photon and laser probes.
%
%
In situations where measurement resources are limited, for example, where the process is rapidly changing or the time bandwidth is constrained, it becomes essential to overcome this precision limit.
Here we devise and demonstrate a scheme for tomography
which exploits
non-classical input states and quantum interferences; unlike previous QPT methods our scheme capitalises upon the possibility to use simultaneously multiple photons per mode. The efficiency---quantified by precision per photon used---scales with larger photon-number input states. Our demonstration uses four-photon states and our results show a substantial reduction of statistical fluctuations compared to traditional QPT methods---in the ideal case one four-photon probe state yields the same amount of statistical information as twelve single probe photons.}

Quantum information \red{protocols} promise \red{new} capabilities for a range of computational, communication and sensing applications. Successful \blue{development} and implementation of all these quantum information protocols rely on efficient techniques to characterise quantum devices. The most widely-used method for this purpose is QPT---in which a mathematical description of a quantum process is reconstructed by \blue{estimating} the probabilities of outcomes for a selection of probe states and measurement settings.  QPT has been demonstrated in a variety of physical systems, including ion traps~\cite{ri-prl-97-220407}, nuclear magnetic resonance~\cite{ch-pra-64-012314}, superconducting circuits~\cite{bi-nphys-6-409} and Nitrogen-vacancy colour centres~\cite{ho-njp-8-33}. 
In the context of photonics, experimental demonstrations have been reported in \red{e.g.} Refs.~\cite{na-spie-4917-13, ob-prl-93-080502, de-pra-67-062307, al-prl-90-193601} for linear-optical systems using few-photon states. \red{(}A continuous-variable version of QPT has also been demonstrated for general optical quantum processes using coherent-state inputs and homodyne measurements~\cite{lo-sci-322-563}.\red{)}

Despite the success of QPT on small systems, there remains scope for improvement. For example, a well-known problem for QPT is the requirement for an exponentially-growing number of measurements for processes on an increasing number of qubits. Many methods have been devised and demonstrated to circumvent this problem, such as efficient state tomography~\cite{cr-ncomm-1-149} and compressed sensing~\cite{gr-prl-105-150401}. In photonics, photon loss and interferometric instability present the main challenges, and a method called ``super-stable tomography'' was recently proposed to address these~\cite{la-arxiv-1208-2868}. \red{Here} we are concerned with improving measurement precision given a fixed number of particles propagating through the unknown process. 
\blue{It has been found that the precision achieved by tomography methods after a fixed number of measurements are dependent on the unknown state or process. This has naturally led to adaptive schemes~\cite{su-pra-85-052107,ma-prl-111-183601} where dynamic measurement settings are used.} Fundamentally, whatever measurement scheme is used---adaptive or non-adaptive---the precision is always limited by the unavoidable statistical fluctuation, \red{where} the ultimate precision limits are dictated by quantum mechanics. 

Our \red{approach} is to exploit quantum interferences to minimize the unwanted fluctuation on the quantum-process measurement statistics by drawing upon techniques from quantum metrology~\cite{gi-nphot-5-222}. \red{Here} we present a quantum-enhanced process tomography protocol which works for arbitrary unitary optical processes on two modes, and experimentally demonstrate measurement precision beyond that achievable with traditional QPT protocols. We first explain the theory of our protocol, and then present the results of our experimental implementation which show evidence for a quantum-enhanced precision compared to the conventional QPT approach. Finally, we discuss \red{generalisations and applications of} our protocols.

The standard procedure for QPT applies repeated state tomography on a set of input states acted on by the process \cite{NielsenChuang,ob-prl-93-080502}.  A full procedure for process tomography commonly assumes that the quantum process corresponds mathematically to completely-positive trace-preserving map \blue{and physically to quantum evolution which can include decoherence or dissipation}. 
If the process acts on a $l$-dimension system, $l^4-l^2$ configurations must be tested~\cite{NielsenChuang}.  
Here we consider the unitary case where there are $l^2-1$ unknown real parameters, encompassing a broad class of optical devices and processes. For the case of two-mode unitaries~($l=2$), there are $3$ real parameters that need to be determined and they correspond to the complex transmission amplitude $a+ib$ and the complex reflection amplitude $c+id$. So the task to estimate the unknown unitary $U$ becomes to determine the values of $a$, $b$, $c$ and $d$ satisfying the unitarity constraint
\begin{equation}
a^2+b^2+c^2+d^2=1.
\label{eq:constraint}
\end{equation}
We assume that the unitary is operating on the polarisation degree of freedom and we denote horizontal~(diagonal, right circular) and vertical~(anti-diagonal, left circular) polarisation by H~(D, R) and V~(A, L), where $\ket{D/A}\!=\!\frac{1}{\sqrt{2}}(\ket{H}\!\pm\!\ket{V})$ and $\ket{R/L}\!=\!\frac{1}{\sqrt{2}}(\ket{H}\!\pm\!i\ket{V})$. By inputting an H~(D, R) photon to the unitary and measuring the output photon in the H/V~(D/A, R/L) basis, the probability of detecting H~(D, R) polarisation at the output is $p_{HV}$~($p_{DA}$, $p_{RL}$), where
\begin{eqnarray}
\label{eq:single-photon-prob}
p_{HV}&=a^2+b^2,\nonumber\\
p_{DA}&=a^2+d^2,\\
p_{RL}&=a^2+c^2. \nonumber
\end{eqnarray}
By using the unitary constraint of Eq.~\ref{eq:constraint}, $U$ can then be estimated by measuring these three probabilities. \blue{There is a discrete set of estimates, which all correspond to the same values for $p_{HV(DA,RL)}$. (A similar situation exists in interferometric phase estimation where typically multiple values of the phase are consistent with a particular set of data.) While the sign of $a$ can always be fixed to positive, the signs of  $b$, $c$ and $d$ need to be resolved. For this we use supplementary standard QPT, using a small and minimal number of probe photons, to provide an initial coarse-grained estimate sufficient to differentiate these alternatives~(See Methods).}


The traditional approach would directly estimate $p_{HV}$, $p_{DA}$ and $p_{RL}$ with single photons by looking at the ratio of detections at the two outputs.
The precision of estimating $p_{HV}$ is $\Delta p_{HV}=\sqrt{p_{HV}(1-p_{HV})/N}$ which scales as $O(N^{-1/2})$, the standard quantum limit~(SQL) for measurement. To go beyond the SQL, our approach uses multi-photon states as probes, determining the three probabilities shown in Eq.~\ref{eq:single-photon-prob}, indirectly from the multi-photon counting statistics.

The multi-photon input state we use is a $N$-photon state split equally between H- and V-polarisation~\cite{ho-prl-71-1355}, $\vert N/2,N/2 \rangle_{HV}$, where $N$ is even. After propagating through the unknown unitary, the state is measured in the H/V basis as in the single-photon case. The probability of detecting $n_H$ H-polarised photons and $n_V$ V-polarised photons at the output is a function of $p_{HV}$ alone. Explicitly, these probabilities are
\begin{equation}
\label{eq:multi-photon-prob}
\mathcal{P}(n_H,n_V,p_{HV})=\frac{n_V!}{n_H!}\left(\mathcal{L}_{(n_H+n_V)/2}^{(n_H-n_V)/2}\left(2p_{HV}-1\right)\right)^2,
\end{equation}
where $\mathcal{L}_{(n_H+n_V)/2}^{(n_H-n_V)/2}$ denotes the standard associated Legendre polynomial~\cite{olver2010nist} with degree $(n_H+n_V)/2$ and order $(n_H-n_V)/2$~(See Supplementary Information). Consequently, $p_{HV}$ can be estimated from the photon-counting data using a maximum-likelihood technique with a precision of $\Delta p_{HV}=\sqrt{p_{HV}(1\!-\!p_{HV})/\left(N(N/2\!+\!1)\right)}$, which scales with $O(N^{-1})$---a quadratic improvement compared to the SQL. For the $N\!=\!4$ case, the outcome probabilities are given in Table~\ref{table:FourProbs}, and the precision of the estimated probability improves by approximately $70\%$. By changing the input state to $\vert N/2,N/2 \rangle_{DA~(RL)}$ and the measurement basis to $D/A~(R/L)$, we can obtain $p_{DA}$~($p_{RL}$) with the same precision as $p_{HV}$. 
It should be noted that it is still not known what kind of quantum advantages can be achieved for the non-unitary processes. Some theoretical results on estimating both unitary and non-unitary processes can be found in References~\cite{ji-ieee-54-5172,kn-qph-1307-0470,cr-qph-1206-0043}.

\begin{table}[t]
\begin{tabular}{cc|cc}
  $(n_{H},n_{V})$ & & & outcome(s) probability \\
  \hline
  $(0,4)$, $(4,0)$ & & & $6p_{HV}^2(1-p_{HV})^2$ \\
  $(1,3)$, $(3,1)$ & & & $6p_{HV}(1-p_{HV})(2p_{HV}-1)^2$ \\
  $(2,2)$ & & & $(6p_{HV}^2-6p_{HV}+1)^2$ \\

\end{tabular}
\caption{The outcome probabilities of four-photon events for the $H/V$ measurement. Analogous expressions hold for the $D/A$ and $R/L$ measurements.}
\label{table:FourProbs}
\end{table}

\begin{figure}[t]
\begin{center}
\includegraphics[width=0.5\textwidth]{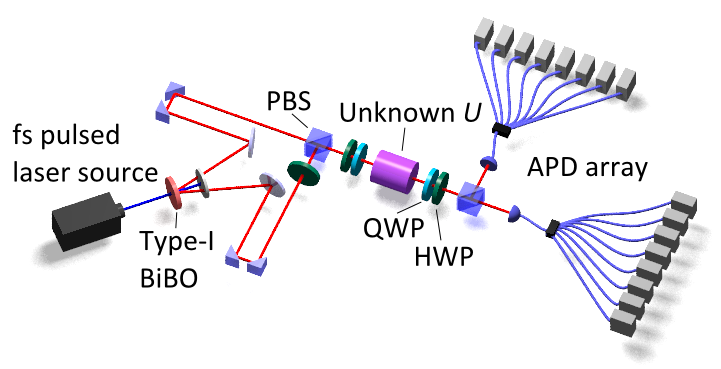}
\caption{\label{fig:thesetup} \textbf{Experimental setup for probing the unknown unitary processes.} An 80MHz pulsed Ti-Sapphire laser centred at 808nm is up-converted to 404nm and then focused onto a BiBO~(Bismuth Borate) crystal, phase-matched for type-I SPDC, creating non-collinear degenerate horizontally-polarised photon pairs at 808nm. After converting one arm to vertical polarisation using a half-waveplate~(HWP, green), the two arms are combined by a polarising beamsplitter~(PBS). The resulting state is passed through the unknown unitary~(purple) and then measured in the H/V basis. Approximate photon-number counting is implemented on each polarisation mode using a 1-8 fan-out array onto avalanche photodiode detectors (APD, grey).  By using different waveplate settings before and after the unknown unitary, the basis for the probe state and measurement can be changed to D/A and R/L. }
\vspace{-7mm}
\label{fig:setup}
\end{center}
\end{figure}

\begin{figure}[t]
\begin{center}
\includegraphics[width=0.5\textwidth]{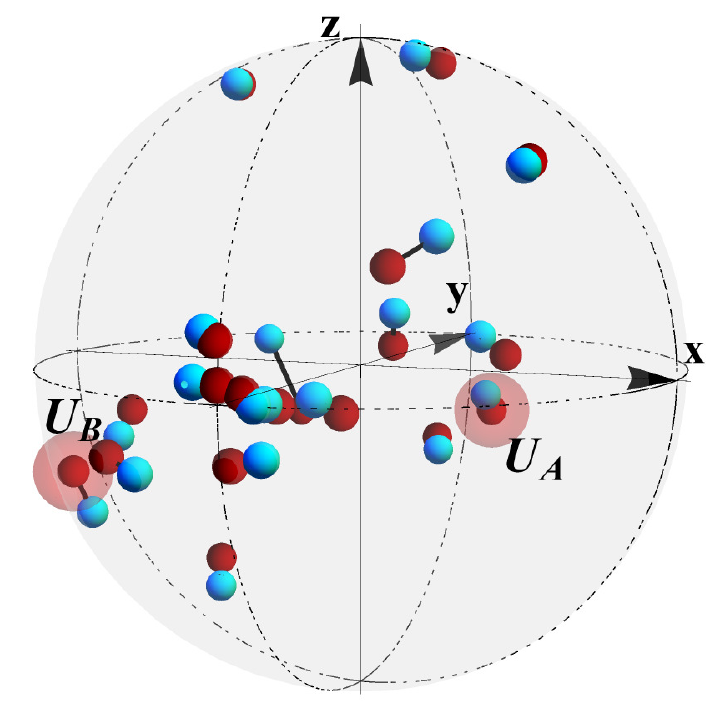}
\caption{\textbf{Visual representation of 19 randomly-selected unitaries~(red) with their experimental reconstructions from four-photon probe states~(cyan).} Each of these unitaries was experimentally realised with three consecutive waveplates~(HWP, QWP, HWP). The four-photon state $\vert 2,2 \rangle_{HV~(DA, RL)}$ is used to probe each unitary and subsequently measured in the H/V~(D/A, R/L) basis. The unitaries are then reconstructed from the resulting photon statistics. The representation of each unitary 
is a point in this unit sphere with coordinates $(x, y, z)$ corresponding to $(-d, c, -b)$.
\blue{The precision of two highlighted unitaries, $U_A$ and $U_B$, is subject to detailed analysis with a large data set in Fig.~\ref{fig:SU(2)_bar_chart}.}}
\label{fig:19-reconstructions}
\end{center}
\end{figure}

\begin{figure*}[t]
\begin{center}
\includegraphics[width=\textwidth]{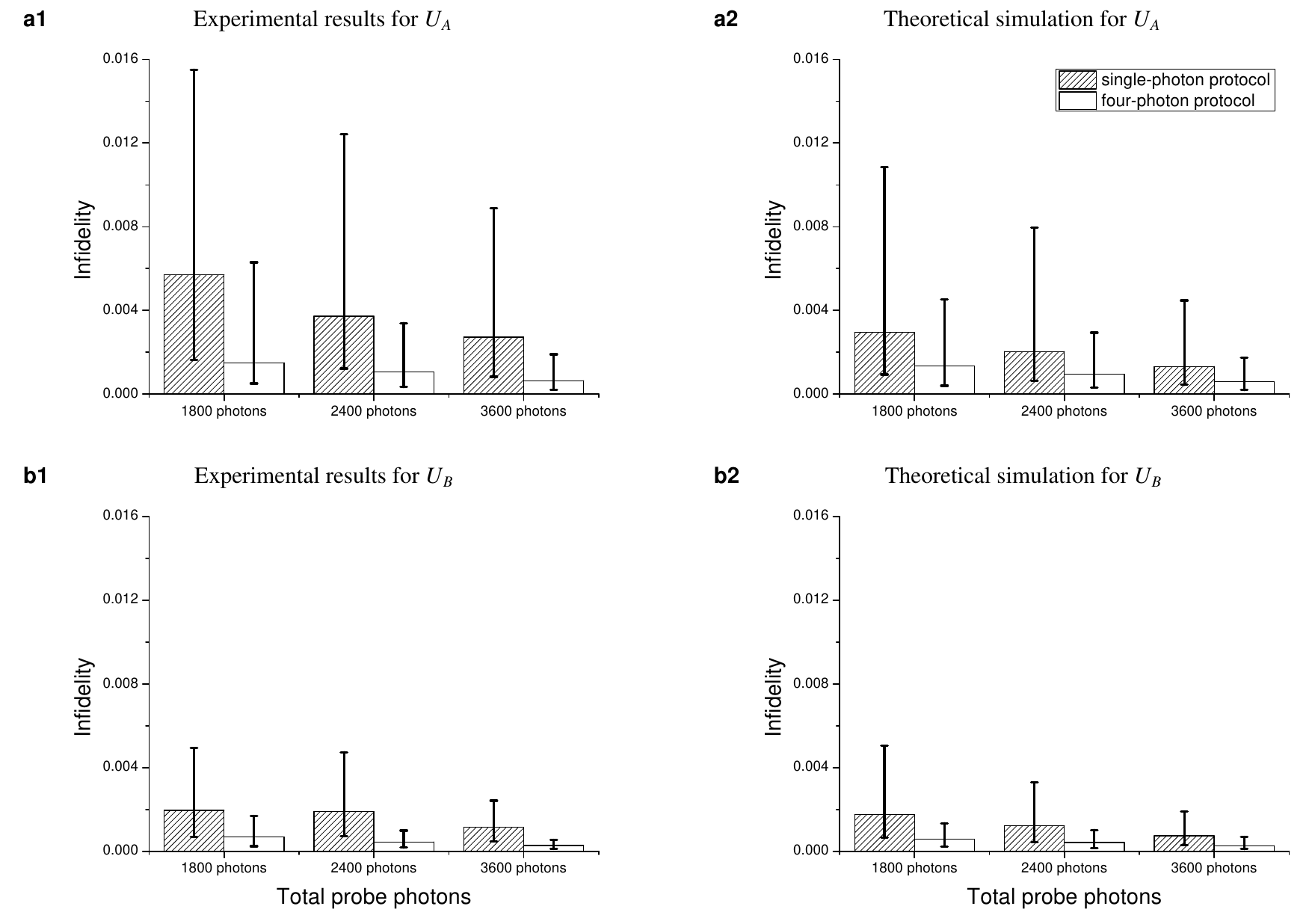}
\vspace{-3 mm}
\caption{\blue{\textbf{Experimental performance of the quantum enhanced tomography protocol using four-photon states.}} In this figure, the standard QPT protocol~(diagonal fill) using single photons and our method using four photons~(unpatterned) are compared, for two unitaries $U_A$ and $U_B$. To make a fair comparison between the two protocols, the same total number of probe photons are assumed (1800, 2400 and 3600 for all measurements together). \textbf{a1} and \textbf{b1} are derived from experimental data for $U_A$ and $U_B$ respectively and \textbf{a2} and \textbf{b2} are based on corresponding theoretical simulations. For each case, the mean infidelity is illustrated, for a series of experimentally-derived estimates $\tilde{U}_A$ or $\tilde{U}_B$, where the infidelity is defined relative to ``central'' estimates $U'_A$ or $U'_B$ derived from the accumulated experimental data. Asymmetric lower- and upper-half error bars are used, reflecting the distribution of infidelity values. The data shows that the four-photon version of the protocol achieves greater accuracy and precision than standard QPT with single photons.}
\label{fig:SU(2)_bar_chart}
\end{center}
\end{figure*}



To demonstrate our scheme, we use four-photon states generated using standard type-I spontaneous parametric down-conversion~(SPDC)~(see Fig.~\ref{fig:setup}). With a certain probability, the SPDC source will produce two photon pairs~(with H polarisation) across the two arms. After rotating one arm to V polarisation using a half-waveplate~(HWP), the two arms are then combined on a polarisation beamsplitter~(PBS) thus producing the desired four-photon state $\vert 2,2 \rangle_{HV}$. The state passes through the unknown unitary and is then separated into the H and V components using a PBS. The photon number at each output is resolved using a fan-out array which couples to eight avalanche photodiodes~(APDs)~(See Methods). We use the measured rates of four-photon outcomes to estimate $p_{HV}$ by using the maximum-likelihood method based on the theoretical probability distributions shown in Table~\ref{table:FourProbs}. By changing the input state and the measurement basis to D/A~(R/L), implemented by waveplates before and after the unitary, we estimate $p_{DA}$~($p_{RL}$) in a similar manner. The experimentally determined $\tilde{p}_{HV}$, $\tilde{p}_{DA}$ and $\tilde{p}_{RL}$ are then used to construct an estimate of the unknown unitary, $\tilde{U}$.  To quantify the discrepancy between $\tilde{U}$ and $U$ we use the process infidelity, defined as $\left(1-{\rm min}\left\vert\bra{\psi}\tilde{U}^{\dag}U\ket{\psi}\right\vert^2\right)$, where the minimum is taken over all single-photon states $\ket{\psi}$.

To provide evidence that the scheme works for any two-mode unitary, we test it using a number of preselected unitaries randomly sampled from the Haar distribution \cite{me-nams-54-592}.  For each of these unitaries $U$, 200 four-photon probes (800 photons in total) are used to construct the estimate $\tilde{U}$.  The process infidelities range from $96.8\%$ to $99.8\%$, with a mean of $98.8\%$ showing near-ideal performance for the scheme.  The relationships between the expected and the experimentally-reconstructed unitaries are represented graphically in Fig.~\ref{fig:19-reconstructions}. This analysis shows qualitatively that our scheme provides high-quality estimates for arbitrary unitaries.

Now we turn to the central feature of the scheme---the ability to exploit multi-photon quantum interference to improve the estimate precision with a fixed input resource---the total number of photons propagating through the unknown unitary. We choose two of these unitaries, \blue{$U_A=\bigl(\begin{smallmatrix}
0.70+0.21i&-0.65-0.20i\\ 0.65-0.20i&0.70-0.21i
\end{smallmatrix} \bigr)$ and $U_B=\bigl(\begin{smallmatrix}
0.29+0.34i&0.33+0.83i\\ -0.33+0.83i&0.29-0.34i
\end{smallmatrix} \bigr)$~(shown in fig.~\ref{fig:19-reconstructions})}, and look at the variation of $\tilde{U}$ over many repetitions of the experiment. Theoretically, both the mean and standard deviation of the process infidelities of $\tilde{U}$ with respect to $U$ will be improved towards zero using our scheme as opposed to the traditional approach using single photons. In practice, we use a SPDC source and post-selection to simulate exact photon-number states, which inevitably result in systematic errors that prevent a fair comparison between the mean infidelities of four-photon and single-photon probe states~\cite{th-prl-107-113603}. To properly quantify the spread of the estimates $\tilde{U}$ from actual data, we use the standard deviation of the process infidelities of $\tilde{U}$ with respect to $U'$, where $U'$ is a ``central'' estimate derived by combining all datasets.  The analysis of the experimental data for $U_A$ and $U_B$ are shown in Fig.~\ref{fig:SU(2)_bar_chart}. As expected, the mean and the standard deviation of the infidelity decrease as the photon number increases for both schemes. More importantly, for every fixed input resource, the standard deviation of the infidelity is reduced for our scheme compared to the traditional approach. \blue{For example, as shown in Fig.~\ref{fig:SU(2)_bar_chart}, using standard QPT one would need approximately 3600 photons overall to obtain the same small infidelity and spread, as achieved by our protocol using only 1800 probe photons.} The experimental results are closely matched to the predictions of the theoretical simulations as described in Fig.~\ref{fig:SU(2)_bar_chart}. In the ideal case, where there is no restriction on the number of probe photons, one four-photon probe state yields the same amount of statistical information as twelve single probe photons. However, our results show a slightly smaller quantum advantage because of practical limitations on the size of the dataset.











The deviations of our experimental results from the theoretical predictions originate from three parts of the experiment---(i) \emph{Input states}: Our scheme assumes perfect Fock states which have fixed photon number. To simulate the Fock states experimentally, we used a SPDC source which inherently contains higher-order terms, temporal distinguishability between photon pairs and spectral distinguishability between the two arms, all of which alter the intended quantum interference; (ii) \emph{Optical components}: There always is imprecision in setting the angles of the manual waveplates, which can be improved by using a motorised bulk-optical system or migrating to an integrated architecture; (iii) \emph{Detection system}: To simulate photon-number resolving detection, we use a 1-to-8 fibre array terminated with 8 APDs on each output. The non-uniform efficiency across the 16 APDs and the non-identical splitting ratio of the fibre arrays result in some bias in the detection system. Despite these limitations, we still see a clear quantum advantage of our four-photon data over the traditional method.

The ability to characterise a quantum process accurately is a basic requirement for demonstrating quantum information techniques, and the method of QPT has been widely deployed. However, it is not well explored how and to what extent it is possible to exceed the precision limit achievable by QPT. From an alternative perspective, process tomography or unitary estimation, can be regarded as a multi-parameter estimation problem. A general linear-optical unitary on $N$ spatial modes can always be decomposed into multiple variable phases and 50:50 beamsplitters~\cite{re-prl-73-58}. Recently, some special families of unitaries using commuting phaseshift operations, with applications in phase imaging~\cite{hu-prl-111-070403} and interferometry in waveguides~\cite{sp-srep-2-862}, have been theoretically investigated.

Our protocol, for the general case of a two-mode unitary, corresponds to estimating three unknown non-commuting phases. Since quantum enhancement is well known for single phase estimation in the field of quantum metrology, it is natural to look for an analogous improvement for this multi-parameter problem. We have drawn upon and adapted the techniques from optical phase estimation and successfully achieved a quantum enhancement in estimating a unitary process. A key strength of this approach is our parametrisation which greatly simplifies the maximum-likelihood procedure. As an aside, we note that there have been several related theoretical investigations which explore how the properties of quantum mechanics, especially quantum entanglement, can improve the precision for abstract unitary estimation~\cite{ka-pra-75-022326,ba-qph-0507073,ac-pra-64-050302,ha-pla-354-183}. However, these papers adopt different methodologies from our work, and offer no explicit mapping onto photonic systems.

Although our scheme is based on an equal number Fock state in two modes $\vert N/2, N/2\rangle$ \red{for the input}, our method can be conveniently extended with only minor modifications for: (1) Unbalanced input states of the form $\vert M, N-M\rangle$, and superpositions of these states with different total photon number; (2) \red{Photon-counting techniques with limited ability to discriminate exact photon number}~\cite{sp-pra-85-023820}. The modifications involve changes to the probability distributions used in the maximum-likelihood estimation procedure, but do not alter \red{the special feature that they always depend on a single parameter}~($p_{HV}$, $p_{DA}$ or $p_{RL}$). As a practical application, the probe state can be the entire state generated by type-I or type-II parametric downconversion, which can be easily created in the laboratory and is tolerant to photon loss~\cite{th-prl-107-113603,ma-qph-1307-4673}; \red{(3) Unknown unitaries on $n$ modes: the $n^2-1$ parameters of such unitaries can be determined by using the same number of input and measurement bases, where for each choice of basis a non-classical multi-photon input state is used.} (4) Incorporating an adaptive method. As with standard QPT, the attainable precision of our protocol is dependent on the unknown unitary. As such, by including an adaptive step---updating the measurement bases---the precision can be further improved.



Our protocol can be directly used for optical communication networks~\cite{ri-nat-484-195}, transferring classical or quantum information, with the purpose of fast and precise characterisation of each link, especially when interferometric stability is required. Our technique also offers a range of applications to the characterisation of optical media and quantum logic gates~\cite{ob-sci-318-1567}, as well as to new types of quantum sensors~\cite{gi-nphot-5-222}. Considering applications outside of photonics, the scheme has a natural geometric interpretation bestowed by the Schwinger representation~\cite{sakurai1985modern}, which provides a one-to-one map from two-mode-$N$-photon states to the $N/2$-spin state space. Here the two-mode unitary operations correspond to physical rotations of a spin system, or equivalently rotations of the reference frame. Consequently, our experiment shows that a quantum advantage is indeed possible for the task of aligning Cartesian reference frames, as predicted in several theoretical works using other protocols~\cite{ba-rmp-79-555}.  A spin implementation of our protocol has practical applications for both gyroscopy and magnetometry.

\vspace{5 mm}
\noindent\textbf{\large{Methods}}

\small{\noindent\textbf{Reconstruction of $U$ from estimated quantities $\tilde{p}_{HV}$, $\tilde{p}_{RL}$, and $\tilde{p}_{DA}$}

\noindent Linear inversion of Eq.s~\ref{eq:constraint} and \ref{eq:single-photon-prob} allows the unknown unitary to be reconstructed from the experimental-derived estimates $\tilde{p}_{HV}$, $\tilde{p}_{RL}$ and $\tilde{p}_{DA}$, 
whenever this leads to nonnegative values for $\tilde{a}^2$, $\tilde{b}^2$, $\tilde{c}^2$, or $\tilde{d}^2$.  This inversion is given explicitly by,
\begin{equation}
\left(
\begin{array}{c}
\tilde{a}^2 \\
\tilde{b}^2 \\
\tilde{c}^2 \\
\tilde{d}^2 \\
\end{array}
\right)
=
\frac{1}{2}
\left(
\begin{array}{cccc}
-1 &  1  &   1  &  1 \\
1  & -1  &  -1  &  1 \\
1  & -1  &   1  & -1 \\
1  &  1  &  -1  & -1
\end{array}
\right)
\left(
\begin{array}{c}
1 \\
\tilde{p}_{DA} \\
\tilde{p}_{RL} \\
\tilde{p}_{HV}
\end{array}
\right),
\end{equation}
and it can be applied whenever the following inequalities are all satisfied:
\begin{eqnarray}
&& \tilde{p}_{DA} + \tilde{p}_{RL} + \tilde{p}_{HV} \geq 1 \nonumber \\
&& \tilde{p}_{DA} + \tilde{p}_{RL} - \tilde{p}_{HV} \leq 1 \nonumber \\
&& \tilde{p}_{DA} - \tilde{p}_{RL} + \tilde{p}_{HV} \leq 1 \nonumber \\
&& -\tilde{p}_{DA} + \tilde{p}_{RL} + \tilde{p}_{HV} \leq 1. \nonumber
\end{eqnarray}
Possible values for the probability estimates define the cube with $0\leq\tilde{p}_{HV},\tilde{p}_{RL},\tilde{p}_{DA}\leq 1$, and the inequalities above determine a tetrahedral subregion which we call the {\it physical region}\blue{---with vertices at (1,0,0), (0,1,0), (0,0,1) and (1,1,1)}.  Outside of the physical region, exactly one of the inequalities fails to hold, and therefore we choose the point closest  to $\left(\tilde{p}_{HV}, \tilde{p}_{RL},\tilde{p}_{DA}\right)$ in the physical region (with respect to the Euclidean metric). Simple expressions for the closest point follow from geometric considerations. As an aside, the maximum-likelihood procedure which is applied to estimate the value of $\tilde{p}_{HV(DA,RL)}$ from data, with four-photon input states, cannot distinguish values $\tilde{p}_{HV(DA,RL)}$ and $1-\tilde{p}_{HV(DA,RL)}$, which are both consistent with measurement results. This is because the probability distributions in Table \ref{table:FourProbs} and Eq. \ref{eq:multi-photon-prob} are symmetric under the mathematical operation $p_{HV}\leftrightarrow1-p_{HV}$ (and similarly for $p_{DA/RL}$). This ambiguity is resolved by the same coarse-grained estimates, obtained with supplementary QPT measurements, which is needed to determine the signs of $b$, $c$ and $d$.

\vspace{5 mm}
\noindent\textbf{\large{Supplementary Information}}
\section{Parameter dependence of the probability distributions for measurements in a fixed basis}

Here we identity what information is obtainable from each measurement in our protocol, for an arbitrary (unitary) linear-optical process $U$ on two modes.  The action of $U$ on the mode operators is given by,
\begin{eqnarray}
\begin{pmatrix}
a_{H}^{\dag \prime } \\
a_{V}^{\dag \prime } \\
\end{pmatrix}
=
\begin{pmatrix}
U a_{H}^{\dag }U^\dag \\
U a_{V}^{\dag }U^\dag \\
\end{pmatrix}
= {\mathcal M}^t
\begin{pmatrix}
a_{H}^{\dag } \\
a_{V}^{\dag } \\
\end{pmatrix}
\end{eqnarray}
where $\mathcal M$ is a unitary two-by-two matrix.  The global phase of $U$ is unmeasurable in our setup and hence we assume
$\mathcal M \in SU(2)$.

$\mathcal M$ also corresponds to the linear transformation by $U$ of an arbitrary single-photon superposition state in the Fock basis,
$\ket{\psi_1} = c_H \ket{1,0}_{HV} \!+\! c_V\ket{0,1}_{HV}$,
so that $\ket{\psi_1} \mapsto U \ket{\psi_1}$ is given by
$
\left(
\begin{smallmatrix}
c_H \\ c_V \\
\end{smallmatrix}
\right)
\mapsto
{\mathcal M}
\left(
\begin{smallmatrix}
c_H \\ c_V \\
\end{smallmatrix}
\right)
$.
We can represent $\ket{\psi_1}$ geometrically on the Bloch sphere with $\ket{0} \equiv \ket{H}$ and $\ket{1} \equiv \ket{V}$ in the usual qubit notation.  $U$ then acts by rotating the Bloch vector of $\ket{\psi_1}$ by an angle $\phi$ around the rotation axis with unit vector $\boldsymbol n$, where
${\mathcal M}=\exp \left[ -i(\phi /2){\bf n} \cdot {\boldsymbol \sigma} \right]$ (${\boldsymbol \sigma}=\left(\sigma_x,\sigma_y,\sigma_z\right)$ denotes the Pauli matrices).
For an arbitrary $N$-photon state,
$\left\vert \psi_N \right\rangle =\sum_{M=0}^{N}c_{M}a_{H}^{\dag M}a_{V}^{\dag N-M}\left\vert \text{vac}\right\rangle $,
$U\left\vert \psi_N \right\rangle =\sum_{M=0}^{N}c_{M}\left( a_{H}^{\dag \prime }\right) ^{M}\left( a_{V}^{\dag \prime }\right) ^{N-M}\left\vert \text{vac}\right\rangle$,
and again the transformation is determined entirely by the coefficients of $\mathcal M$.

Next we look at the general form of the probability distributions for measuring $n_{H(V)}$ horizontally (vertically)-polarized photons at the output, given state
$\ket{M,N-M}_{HV}$ at the input, with notation ${\mathcal P}_{HV}(n_H,n_V)=\left\vert \bra{n_H,n_V}_{HV} U \ket{M,N-M}_{HV} \right\vert ^2$.  ($M=N/2$ in the main text.)
We can use an Euler-angle decomposition to write $\mathcal M$ as a sequence of rotations on the Bloch sphere about the $y$ and $z$ axes:
$M=\exp \left[ -i(\psi /2)\sigma _{z}\right] \exp \left[ -i(\theta /2)\sigma _{y}\right] \exp \left[ -i(\zeta /2)\sigma _{z}\right]$. The $z$-axis rotations generate phases which do not affect the value of ${\mathcal P}_{HV}(n_H,n_V)$, which therefore depends only on the y-axis rotation with angle $\theta$.  As in the main text, we can use $p_{HV}$ to parameterize ${\mathcal P}_{HV}(n_H,n_V)$, and $p_{HV} = \cos^2(\theta/2).$
The probability distributions are given explicitly by rotational Wigner $d$-matrices as follows,
\begin{equation}
{\mathcal P}_{HV}(n_H,n_V,p_{HV})=
\left [ d^\frac{N}{2}_{{\frac{n_H}{2}-\frac{n_V}{2}},M-\frac{N}{2}}
\left(2\arccos \sqrt{p_{HV}} \right) \right ]^2
\!\!.
\end{equation}
(see Ref.~\cite{sakurai1985modern} for a derivation of the $d$-matrices).  For the case $M=N/2$, ${\mathcal P}_{HV}(n_H,n_V,p_{HV})$ can be reexpressed using the associated Legendre polynomials as given explicitly in Eq.~3 in the main text.

\section{Performance of our protocol with increasing number of probe photons}
\begin{figure}[t]
\includegraphics[width=0.9\linewidth]{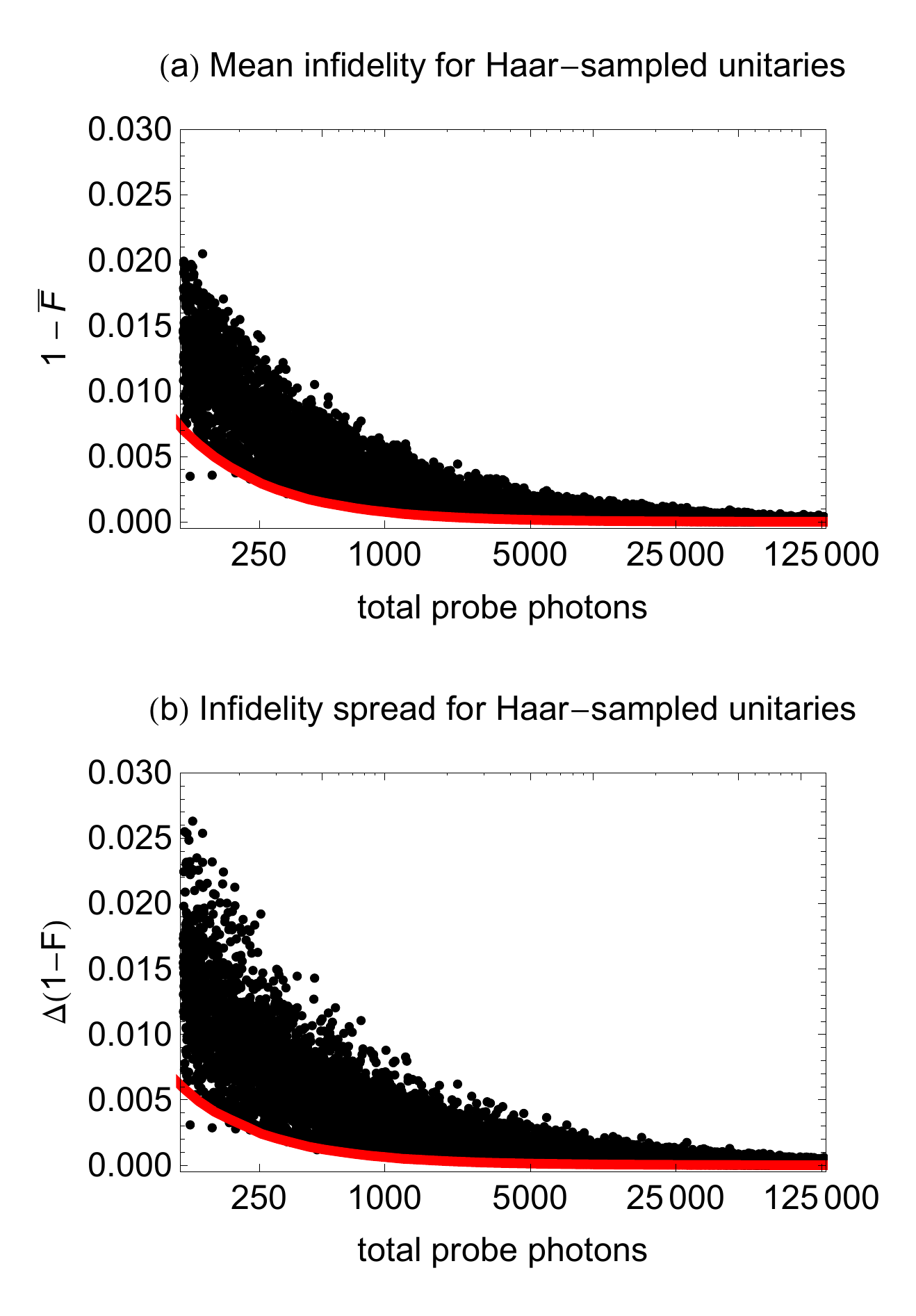}
\caption{\label{fig:protocolgeneralperformance}
These plots display the results of a simulation of our protocol for 10,000 randomly-chosen unitary operations (using the Haar-measure uniform distribution for U(2)).  Each point corresponds to one unitary, and a randomly-chosen number of total probe photons. In every case, we assume four-photon input states: $\vert 2,2\rangle_{HV(DA,RL)}$.  The red lines show the perfoprmance for the unitaries at the centre of the physical region: these have $a=\pm1/2$, $b=\pm1/2$, $c=\pm1/2$ and $d=\pm1/2$.}
\end{figure}

Here we present the performance of our protocol for a variety of unknown $U$ and varying numbers of probe photons.  To quantify the closeness of an estimate of $\tilde{U}$ to $U$ itself we use the process infidelity $1-F$, defined as in the main text as $\left(1-{\rm min}\vert\langle \psi \vert \tilde {U}^\dag U \vert \psi \rangle\vert^2\right)$, where the minimization is over single-photon states.  The minimization can been done analytically, giving $1-F=1-\left(a\tilde{a}+b\tilde{b}+c\tilde{c}+d\tilde{d}\right)^2$, where $a+ib$ and $c+id$ are the transmission and reflection amplitudes for $U$, and $\tilde{a}+i\tilde{b}$ and $\tilde{c}+i\tilde{d}$ are the corresponding estimated values.

Fig.~\ref{fig:protocolgeneralperformance} shows the performance of our protocol for randomly-chosen $U$ using four-photon input states.  The choice of $U$ affects both the sensitivity of each of the measurements used in the protocol, as well as the proportion of estimates (by linear inversion) that lie in the physical region; both of these factors affect the mean and spread of the infidelity (for a fixed total number of probe photons).  Fig.~\ref{fig:su2-3vs21} compares the results of a simulation of the performance of our protocol with unitaries $U_A$ and $U_B$ for single and four-photon inputs states, showing how the mean and spread of the infidelity converge to 0 as the number of probe photons increases.  We can observe that the errors for estimating each unitary are always less using the four-photon input states in our protocol than when single-photon input states are used (for the same number of probe photons).

\begin{figure}[t]
\centering
\includegraphics[width=0.6\linewidth]{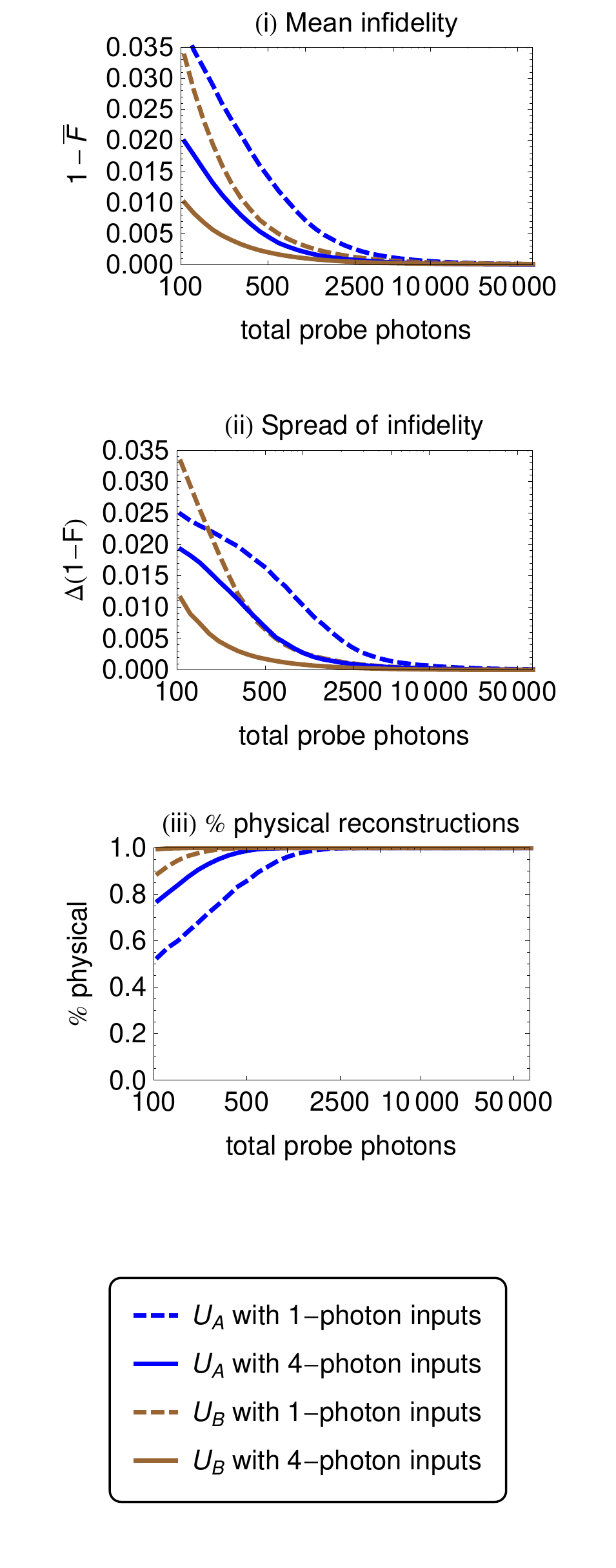}
\caption{\label{fig:su2-3vs21} The plots show the results of a simulation of the performance of our protocol for unitaries $U_A$ and $U_B$, defined by
comparing the cases of single-photon inputs states
$\vert 1,0\rangle_{HV(DA,RL)}$
and four-photon input states,
$\vert 2,2\rangle_{HV(DA,RL)}$.}
\end{figure}

\vspace{5 mm}

\noindent\textbf{Acknowledgements} The authors are grateful for financial support from EPSRC, ERC, NSQI, NRF (SG) and MOE (SG).
JCFM is supported by a Leverhulme Trust Early-Career Fellowship.
JLOB acknowledges a Royal Society Wolfson Merit Award and a RAE Chair in Emerging Technologies.
We thank Tomek Paterek, Peter Turner and Sai Vinjanampathy for helpful discussions.

\end{document}